\documentclass[fleqn,usenatbib]{mnras}

\usepackage{newtxtext,newtxmath}
\usepackage[T1]{fontenc}
\usepackage{ae,aecompl}

\usepackage{graphicx}	% Including figure files
\usepackage{amsmath}	% Advanced maths commands
\usepackage{amssymb}	% Extra maths symbols
\usepackage{array}      % For wrapping long text in table
\usepackage{multirow}
\title[Update on Testing the Isotropy]{Update on Testing the Isotropy of the Properties of Gamma-Ray Bursts}

\author[Jakub \v{R}\'{\i}pa \& Arman Shafieloo]{
Jakub \v{R}\'{\i}pa,$^{1,2,3}$
Arman Shafieloo,$^{4,5}$\thanks{E-mail: shafieloo@kasi.re.kr}
\\
$^1$Astronomical Institute of Charles University, V Hole\v{s}ovi\v{c}k\'ach 2, CZ-180 00 Prague 8, Czech Republic\\
$^2$MTA-E\"otv\"os University Lend\"ulet Hot Universe Research Group, P\'azm\'any P\'eter s\'et\'any 1/A, Budapest, 1117, Hungary\\
$^3$Institute of Physics, E\"otv\"os University, P\'azm\'any P\'eter s\'et\'any 1/A, Budapest, 1117, Hungary\\
$^4$Korea Astronomy and Space Science Institute, Daejeon 305-348, Korea\\
$^5$University of Science and Technology, Daejeon 34113, Korea\\
}

\date{Accepted XXX. Received YYY; in original form ZZZ}

\pubyear{2019}

\begin{document}
\label{firstpage}
\pagerange{\pageref{firstpage}--\pageref{lastpage}}
\maketitle

\begin{abstract}
Previously we proposed a novel method to inspect the isotropy of the properties of gamma-ray bursts (GRBs) such as their duration, fluences and peak fluxes at various energy bands and different time scales, complementary to existing studies of spatial distribution of GRBs by other authors. The method was then applied on the {\em Fermi} GBM Burst Catalog containing 1591 GRBs and except one particular direction where we noticed some hints of violation from statistical isotropy, the rest of the data showed consistency with isotropy. In this work we apply our method with some minor modifications to the updated {\em Fermi}/GBM data sample containing 2266 GRBs, thus $\sim 40$\,\% larger. We also test two other major GRB catalogs, the BATSE Current GRB Catalog of the {\em CGRO} satellite containing $\sim 2000$ bursts and the {\em Swift}/BAT Gamma-Ray Burst Catalog containing $\sim 1200$ bursts. The new results using the updated data are consistent with our previous findings and no statistically significant anisotropic feature in the observed properties of these samples of all GRBs is found.
\end{abstract}


%%%%%%%%%%%%%%%%%%%%%%%%%%%%%%%%%%%%%%%%%%%%%%%%%%

%%%%%%%%%%%%%%%%% BODY OF PAPER %%%%%%%%%%%%%%%%%%

\section{Introduction}
\label{sec:intro}

The sky distribution of Gamma-Ray Bursts (GRBs) \citep{pir04,mes06,ved09,kou12,gom12,kum15,dai17,wil17} has been tested intensively and the early works claimed that they were distributed isotropically \citep{mee92,bri96,teg96}. As more observational data increased and various methods were applied, it was claimed that the group of short GRBs ($T_{90}<2$\,s) \citep{bal03}, which originate in mergers of compact objects such as neutron stars \citep{pac86,eic89,ber14,abb17a,abb17b,abb17c}, was distributed anisotropically \citep{bal98,bal99,mag03,vav08,tar17}, where the $T_{90}$ is the duration during which 90\,\% of the detected counts from a GRB is accumulated \citep{kou93}. However, several other works claimed different results \citep{mes00a,mes00b,lit01,ber08,ver10,ukw16}. The works which analyzed GRBs of duration 2\,s $\lesssim T_{90} \lesssim$ 10\,s found that also these bursts were distributed anisotropically \citep{mes00a,mes00b,lit01,vav08,ver10}. However, see works by \citet{mes03,ukw16} which came to different conclusion. Anisotropical distribution on the sky was also proclaimed for very short GRBs ($T_{90} \leq 100$\,ms) \citep{cli05,ukw16}. On the other hand, the group of long GRBs ($T_{90}>2$\,s), which are associated with the collapses of massive stars \citep{fru06,woo06}, were found to be distributed isotropically \citep{bal98,bal99,mes00a,mes00b,mag03,vav08,tar17,ukw16}, however see \citet{mes03}, which came to a different conclusion. Papers \citet{mes09a,mes09b,mes17,mes18} summarize these efforts.

\citet{hor14,hor15} studied the spatial distribution of GRBs with known redshift and they concluded that they found a statistically significant clustering at the redshift range of $1.6 < z \leq 2.1$ of the size of about 2\,000--3\,000\,Mpc. However, the work \citet{ukw16} also analyzed GRBs with measured redshift concluded that an evidence of a significant clustering was not found. Implications of the results obtained by \citet{hor14} on the Cosmological Principle are discussed by \citet{li15}. \citet{bal15,bal18} found an over-density of GRBs in the redshift range of $0.78 < z <0.86$ (comoving distance of 2\,770 Mpc), which forms a ring-like shape displayed by 9 GRBs with a diameter of 1\,720 Mpc. Other recent works which study the spatial distribution of GRBs with known redshift were done by \citet{rai10,shi17}.

The isotropy or homogeneity of the Universe has been intensively analyzed also using various observations other than GRBs, for example: supernovae Ia \citep{col11,fei13,app15,jav15}, galaxies and clusters of galaxies \citep{fri05a,fri05b,got05,hog05,kas08,scr12,fer14,app14,kro15,jav17,sok18}, active galactic nuclei \citep{tiw18}, HI line sources \citep{rub07,avi18}, groups of quasars \citep{clo13,nad13}, cosmic microwave background \citep{hin96,pla14,pla16}.

In our previous work \citet{rip17} we proposed a new method to test the isotropy of the observed properties of GRBs such as their duration, fluences and peak fluxes at various energy bands and different time scales. This approach is novel because all the previous studies based on the GRB data tested only the distribution of the number densities of GRBs, but not their observed properties. In this paper we apply almost the same method, but with slight improvement of its sensitivity, to the considerably updated {\em Fermi} / Gamma-ray Burst Monitor (GBM) data sample, which is larger by $\sim 40$\,\%. Herein we also test the Burst And Transient Source Experiment (BATSE) Current GRB Catalog of the {\em Compton Gamma Ray Observatory (CGRO)} satellite containing more than 2000 GRBs and the {\em Swift} / Burst Alert Telescope (BAT) Gamma-Ray Burst Catalog containing more than 1200 GRBs.

The paper is organized as follows. Section~\ref{sec:data_samples} describes the used data samples. Section~\ref{sec:method} details the used methodology. The results are presented in Section~\ref{sec:results} and discussed in Section~\ref{sec:discussion}. The last Section~\ref{sec:conclusion} summarizes our conclusions.

\section{Data samples}
\label{sec:data_samples}

Three catalogs are examined in this work. The first one is the database of the GBM instrument \citep{mee09} of the {\em Fermi} satellite\footnote{http://fermi.gsfc.nasa.gov/} \citep{atw94}. Particularly, we employ the FERMIGBRST - Fermi GBM Burst Catalog\footnote{https://heasarc.gsfc.nasa.gov/W3Browse/fermi/\\fermigbrst.html}
\citep{gol12,pac12,gru14,vki14,bha16}, which is being constantly updated and which is one of the most complete burst catalog to date.

The second one is the database of the BATSE instrument \citep{fis85} of the CGRO\footnote{https://heasarc.gsfc.nasa.gov/docs/cgro/cgro/} \citep{geh93}. Particularly, we employ the BATSE Current Gamma-Ray Burst Catalog\footnote{https://gammaray.nsstc.nasa.gov/batse/grb/catalog/current/\\index.html}. For details see also the previous BATSE catalogs: 1B \citep{fis94}, 3B \citep{mee96}, 4B \citep{mee98}, and 4Br \citep{pac99}.

The third one is the database of the BAT instrument \citep{bar05} of the {\em Neil Gehrels Swift Observatory}\footnote{https://swift.gsfc.nasa.gov} ({\em Swift}) \citep{geh04}. Particularly, we employ the {\em Swift}/BAT Gamma-Ray Burst Catalog\footnote{https://swift.gsfc.nasa.gov/results/batgrbcat/index.html}, which is being constantly updated.

\subsection{{\em Fermi}/GBM sample}
\label{sec:data_gbm}

Compared to our previous work \citet{rip17}, which used {\em Fermi}/GBM sample containing 1591 GRBs, in this article we use updated sample which contains $\sim 40$\,\% larger number of GRBs. In this updated sample there is 2271 GRBs with the first one detected on 2008/07/14 and the last one detected on 2018/02/25. The following observables from the catalog are applied in our study:

\begin{itemize}

\item \parbox[t]{0.92\columnwidth}{Galactic longitude $l$ ($^{\circ}$), in the catalog denoted as {\em LII}.}

\item \parbox[t]{0.92\columnwidth}{Galactic latitude $b$ ($^{\circ}$), in the catalog denoted as {\em BII}.}

\item \parbox[t]{0.92\columnwidth}{GRB duration $T_{90}$ (s) measured in the energy range of 50--300\,keV.}

\item \parbox[t]{0.92\columnwidth}{Photon peak fluxes $F_{64}$, $F_{256}$, and $F_{1024}$ (ph\,cm$^{-2}$\,s$^{-1}$) on the 64-ms, 256-ms, and 1024-ms timescales in the energy range of 10--1000\,keV and in the catalog denoted as {\em Flux\_64}, {\em Flux\_256}, and {\em Flux\_1024}, respectively.}

\item \parbox[t]{0.92\columnwidth}{Photon peak fluxes $F_{64\mathrm{,B}}$, $F_{256\mathrm{,B}}$, and $F_{1024\mathrm{,B}}$ (ph\,cm$^{-2}$\,s$^{-1}$) on the 64-ms, 256-ms, and 1024-ms timescales in the energy range of 50--300\,keV and in the catalog denoted as {\em Flux\_BATSE\_64}, {\em Flux\_BATSE\_256}, and {\em Flux\_BATSE\_1024}, respectively.}

\item \parbox[t]{0.92\columnwidth}{Fluence $S$ (erg\,cm$^{-2}$) is the time integrated flux over the whole duration of a burst in the energy range of 10--1000\,keV and in the catalog denoted as {\em Fluence}.}

\item \parbox[t]{0.92\columnwidth}{Fluence $S_{\mathrm{B}}$ (erg\,cm$^{-2}$) is the time integrated flux over the whole duration of a burst in the energy range of 50--300\,keV and in the catalog denoted as {\em Fluence\_BATSE}.}

\end{itemize}

Five GRBs have the $T_{90}$ duration measured in a different energy range than the nominal range 50--300\,keV or they have fluence and peak fluxes measured in a different energy range than the nominal one 10--1000\,keV, therefore we removed those five GRBs from our data sample. For all remaining 2266 GRBs the galactic coordinates, $T_{90}$ duration, all peak fluxes and fluences are measured, hence all of them (regardless of their duration, spectral properties, luminosity or measured redshift) define the whole {\em Fermi}/GBM sample and are used in our analysis.

\subsection{{\em CGRO}/BATSE sample}
\label{sec:data_batse}
The BATSE Current Gamma-Ray Burst Catalog contains 2702 GRBs with the first burst on 1991/04/21, the last one on 2000/05/26. The following observables from the catalog are applied in our study:

\begin{itemize}

\item \parbox[t]{0.92\columnwidth}{Galactic longitude $l$ ($^{\circ}$), in the catalog denoted as {\em LII}.}

\item \parbox[t]{0.92\columnwidth}{Galactic latitude $b$ ($^{\circ}$), in the catalog denoted as {\em BII}.}

\item \parbox[t]{0.92\columnwidth}{GRB duration $T_{90}$ (s).}

\item \parbox[t]{0.92\columnwidth}{Photon peak fluxes $F_{64\mathrm{,B}}$, $F_{256\mathrm{,B}}$, and $F_{1024\mathrm{,B}}$ (ph\,cm$^{-2}$\,s$^{-1}$) on the 64-ms, 256-ms, and 1024-ms timescales in the energy range of 50--300\,keV and in the catalog denoted as {\em Flux\_64}, {\em Flux\_256}, and {\em Flux\_1024}, respectively.}

\item \parbox[t]{0.92\columnwidth}{Fluence $S_1$ (erg\,cm$^{-2}$) is the time integrated flux over the duration of the burst in the energy range of 20--50\,keV and in the catalog denoted as {\em Fluence\_1}.}

\item \parbox[t]{0.92\columnwidth}{Fluence $S_2$ (erg\,cm$^{-2}$) is the time integrated flux over the duration of the burst in the energy range of 50--100\,keV and in the catalog denoted as {\em Fluence\_2}.}

\item \parbox[t]{0.92\columnwidth}{Fluence $S_3$ (erg\,cm$^{-2}$) is the time integrated flux over the duration of the burst in the energy range of 100--300\,keV and in the catalog denoted as {\em Fluence\_3}.}

\item \parbox[t]{0.92\columnwidth}{Fluence $S_4$ (erg\,cm$^{-2}$) is the time integrated flux over the duration of the burst at energy >300\,keV and in the catalog denoted as {\em Fluence\_4}.}

\end{itemize}

For the sake of completeness, it should be mentioned that not all of the above-mentioned observables are independent \citep{bag98,bag09,bor06}. The duration $T_{50}$ is not investigated in this work because it strongly correlates with commonly used $T_{90}$.

All 2702 GRBs in the catalog have measured galactic coordinates $l$ and $b$. The number of GRBs in this catalog with a measured given observable is following: $T_{90}$ (2037), $F_{64\mathrm{,B}}$ (2132), $F_{256\mathrm{,B}}$ (2132), $F_{1024\mathrm{,B}}$ (2132), $S_1$ (2100), $S_2$ (2118), $S_3$ (2127), and $S_4$ (1752). If an observable is measured then it is included in our analysis, therefore these are the sizes of the whole {\em CGRO}/BATSE samples (no restriction is put on the bursts' duration, spectral properties, luminosity or measured redshift).

\subsection{{\em Swift}/BAT sample}
\label{sec:data_bat}

The data sample of the {\em Swift}/BAT Gamma-Ray Burst Catalog, which we use in this work, contains 1223 GRBs with the first burst on 2004/12/17, the last one on 2018/05/14. The following GRB observables from the catalog are applied in our analysis:

\begin{itemize}

\item \parbox[t]{0.92\columnwidth}{RA from the BAT refined position (J2000, deg), in the catalog denoted as {\em RA\_ground}.}

\item \parbox[t]{0.92\columnwidth}{DEC from the BAT refined position (J2000, deg), in the catalog denoted as {\em DEC\_ground}. We converted RA and DEC to the galactic longitude $l$ ($^{\circ}$) and latitude $b$ ($^{\circ}$).}

\item \parbox[t]{0.92\columnwidth}{GRB duration $T_{90}$ (s).}

\item \parbox[t]{0.92\columnwidth}{1-s peak energy fluxes $F_1$, $F_2$, $F_3$, $F_4$, $F_5$, $F_6$, and $F_7$ (erg\,cm$^{-2}$\,s$^{-1}$) in the energy ranges of 15--25\,keV, 25--50\,keV, 50--100\,keV, 100--150\,keV, 100--350\,keV, 15--150\,keV and, 15--350\,keV, respectively.}

\item \parbox[t]{0.92\columnwidth}{Fluences $S_1$, $S_2$, $S_3$, $S_4$, $S_5$, $S_6$, $S_7$ (erg\,cm$^{-2}$) which are the time integrated fluxes over the duration of the whole burst $T_{100}$ in the energy ranges of 15--25\,keV, 25--50\,keV, 50--100\,keV, 100--150\,keV, 100--350\,keV, 15--150\,keV and, 15--350\,keV, respectively.}

\end{itemize}

Two spectral models, used for the calculation of the energy fluxes and fluences, are given in the catalog: a simple power-law model, and a cutoff power-law model. The best-fit model was selected based on the record in the catalog. If the information about the best-fit model was not available then we selected the one with higher null probability of the model as recorded in the catalog.

Eleven events from 1223 GRBs in the used data sample are not localized, therefore we removed them from the analysis. One event GRB140716A was triggered twice (trigger numbers 604793 and 604792). In order to avoid any mistake related to the overall duration $T_{90}$, fluxes and fluences, we removed both triggers from the data sample. Therefore the remaining number of GRBs with determined position in the sample is 1210.

From this number of 1210 GRBs the number of bursts with measured a given observable is following: $T_{90}$ (1197); $F_1$, $F_2$ and $F_4$--$F_7$ (1143); $F_3$ (1142); $S_1$--$S_7$ (1177) and these are the sizes of the whole {\em Swift}/BAT data samples (no other restrictions were applied).

\section{Method}
\label{sec:method}

Almost the same method, as proposed by \citet{rip17}, is applied in this work. We refer the reader to that paper to see details of the methodology. Here the method is described only briefly.

Instead of examining the number density of GRBs on the sky, as most of the previous studies did, the key idea of this method is to analyze the sky distribution of the properties of GRBs. Therefore our method do not test the spatial distribution of GRBs. Since our method tests solely the isotropy of the observed properties of GRBs, no conclusion is drawn about their spatial distribution. For example, the over-densities of GRBs claimed by \citet{hor14,hor15}, i.e. the Hercules Corona Borealis Great Wall or the Giant GRB Ring claimed by \citet{bal15,bal18} or the anisotropic distribution of very short GRBs claimed by \citet{cli05,ukw16} cannot be confirmed or rejected using this method.

In this article the method was slightly improved in order to increase the sensitivity in the performed statistical tests. Contrary to the original method, where the distributions of a given measured GRB property from large number of randomly spread patches on the sky were compared with a distribution of the same GRB property for the whole sky, herein we compare the distributions from the patches with the distributions obtained from the complement area of that patches. This methodology is based on the principles of the crossing statistic applied in different aspects in cosmology \citep{sha11,col11,sha12a,sha12b,akr14}.

Four different test statistics are applied to measure the differences between the distributions for a random patch and its complement. In order to infer the significance of potential anisotropies the obtained distributions of the test statistics derived from the measured data are compared with the distributions of the test statistics for randomly shuffled data. This comparison of the measured data with the randomly shuffled once means in essence comparison of the measured data to the isotropically distributed hypothetical sample and thus it is a principal step in this procedure. The null hypothesis is that observed properties of GRBs are distributed isotropically and we are testing against this null hypothesis.

Shortly the main steps of the used method are following (for details see \citet{rip17}). 
A thousand of randomly distributed patches on the sky of a fixed radius $r$ is generated and afterwards their positions are kept fixed. Let $G_\mathrm{q}(x)$ be the empirical distribution function of a given tested observable $x$ of GRBs (e.g. $x=T_{90}$, $S$, $F$, etc.) in a given patch (the number of GRBs in the patch is $q$ and it varies patch to patch). Let $F_\mathrm{p}(x)$ be the empirical distribution function of the same tested GRB observable $x$ in the complement area on the sky of the same patch (the number of GRBs in that complement area is $p$ and it naturally also varies patch to patch).

Next, for each patch compare $G_\mathrm{q}(x)$ and $F_\mathrm{p}(x)$ by calculating four test statistics $\xi=D$, $V$, $AD$, or $\chi^2$, where $D$ is the statistic of the two-sample Kolmogo\-rov--Smirnov (K--S) test \citep{kol33,smi39,pre07}, $V$ is the statistic of the two-sample Kuiper test \citep{kui60,pre07}, $AD$ is the statistic of the two-sample Anderson--Darling (A--D) test \citep{and52,dar57,pet76}, and $\chi^2$ is the statistic of the two-sample Chi-square test \citep{pre07}. In case of $\chi^2$ the frequencies in the binned data (with 10 number of bins) of the two samples (containing logarithmic values $\log x$) are compared instead of the comparison of the empirical distribution functions. For 1\,000 patches one obtains a distribution of 1\,000 values of $\xi^\mathrm{m}$ (superscript $m$ denotes that the quantity is related to the actual measured data sample).

In the next step the measured data sample is randomly shuffled $n=1000$ times, i.e. the values $x_{\mathrm{i}}$ of each measurement are randomly shuffled, however the measured positions of GRBs $\{l_{\mathrm{i}}$, $b_{\mathrm{i}}\}$ are kept fixed. For each patch on the sky the test statistic $\xi^\mathrm{s}$ is calculated comparing the distribution $G_\mathrm{q}^\mathrm{s}(x)$ of the shuffled data for GRBs in the given patch and the distribution $F_\mathrm{p}^\mathrm{s}(x)$ for the patch's complement area (superscript $s$ denotes that the quantity is related to the randomly shuffled data). For each statistic $\xi$ we derive the limiting values $\xi^\mathrm{s}_{5}$, $\xi^\mathrm{s}_{1}$, and $\xi^\mathrm{s}_{0}$ which delimit the highest 5\,\%, 1\,\%, and 0\,\% of all $\xi^\mathrm{s}$ values from all patches in all randomly shuffled data, respectively. The value $\xi^\mathrm{s}_0$ is the maximum of the $\xi^\mathrm{s}$ values.

Next, the distributions of a given statistic $\xi$ for the measured data
and for all data shufflings are compared. Let count the number of patches $N^\mathrm{m}_\mathrm{i}$ in the measured data for which $\xi^\mathrm{m}>\xi^\mathrm{s}_\mathrm{i}$. The mean number of patches $\overline{N^\mathrm{s}_\mathrm{i}}$ in the randomly shuffled data for which $\xi^\mathrm{s}>\xi^\mathrm{s}_\mathrm{i}$ is $\overline{N^\mathrm{s}_\mathrm{i}}$ = 50 and 10 for i = 5 and 1, respectively. If $N^\mathrm{m}_\mathrm{i}$ is found to be much higher than the average numbers $\overline{N^\mathrm{s}_\mathrm{i}}$ then it could indicate anisotropy in the measured data. One can calculate the probability (significance) $P^\mathrm{N}_\mathrm{i}$ of finding at least $N^\mathrm{m}_\mathrm{i}$ number of patches with $\xi^\mathrm{s}>\xi^\mathrm{s}_\mathrm{i}$ in the randomly shuffled data as well.

This analysis is performed for several patch radii $r=20^\circ$, $30^\circ$, $40^\circ$, $50^\circ$, $60^\circ$, for all tested GRB observables in our data samples and for all four test statistics $\xi=AD$, $D$, $V$, and $\chi^2$. Various patch radii are chosen to make the technique sensitive to potential underlying structures of various sizes. The variations in the GRB properties in a patch is limited by the counting statistics and can be reduced by increasing the patch size.

The routines {\em KSTWO} and {\em KUIPERTWO} of the IDL\footnote{http://www.harrisgeospatial.com/ProductsandSolutions/\\
GeospatialProducts/IDL.aspx} Astronomy Users library\footnote{http://idlastro.gsfc.nasa.gov/} \citep{lan93} for calculation of the $D$ and $V$ statistics were used. For calculation of the A--D statistic, our own code, written based on the {\em adk\_1.0-2} package \footnote{https://cran.r-project.org/src/contrib/Archive/adk/} \citep{adk} of the R software\footnote{https://www.r-project.org} \citep{rsoft}, was used.

\section{Results}
\label{sec:results}
This section covers the results obtained for all examined data samples.

\subsection{Results - {\em Fermi}/GBM}
\label{sec:results_gbm}

Tables~\ref{tab:results_gbm_ks}--\ref{tab:results_gbm_chi} summarize the results of all tested fluxes, fluences and duration of GRBs in the {\em Fermi}/GBM sample using the Kolmogorov--Smirnov statistic $D$, Kuiper $V$, Anderson--Darling $AD$, and Chi-square $\chi^2$ statistics, respectively.

For any of the test statistic, any radius nor any tested GRB observable, the probabilities $P^\mathrm{N}_\mathrm{i}$ never decreased below 5\,\%. Also there is no case of a patch, of a given radius and for a given tested GRB observable and given test statistic $\xi$ in the measured data which gives $\xi^\mathrm{m}$ higher than the highest $\xi^\mathrm{s}_{0}$ for patches of all random data shufflings.

The results applied on the {\em Fermi}/GBM sample are fully consistent with isotropy confirming our previous findings.

\subsection{Results - {\em CGRO}/BATSE}
\label{sec:results_batse}

Tables~\ref{tab:results_batse_ks}--\ref{tab:results_batse_chi} summarize the results of all tested fluxes, fluences and duration of GRBs in the {\em CGRO}/BATSE sample using the Kolmogorov--Smirnov statistic $D$, Kuiper $V$, Anderson--Darling $AD$, and Chi-square $\chi^2$ statistics, respectively.

\begin{figure}
\includegraphics[width=\columnwidth]{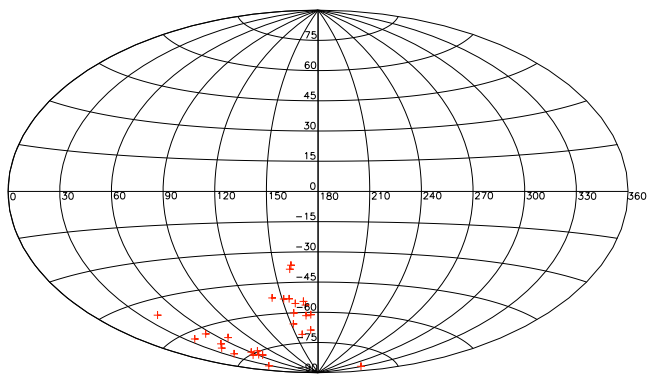}
\caption{Plotted are the patch centers on the sky in Galactic Coordinates (Aitoff projection), for which the $\chi^\mathrm{2\,m}$ statistic, for the measured data, is higher than $\chi^\mathrm{2\,s}_1$ obtained from the randomly shuffled data and the significance $P^\mathrm{N}_1$ is below 5\,\%. The tested GRB property is fluences $S_1$ of the {\em CGRO}/BATSE data sample and the patch radii are $r=20^\circ$. This is the only case in the {\em CGRO}/BATSE sample, for which the significance $P^\mathrm{N}_\mathrm{i}$ is below 5\,\%.}
\label{fig:results_batse}
\end{figure}

There was only one case from all the applied statistical tests when the probability $P^\mathrm{N}_\mathrm{i}$ decreased below 5\,\%. It was for the $\chi^\mathrm{2}$ statistic, fluences $S_1$, patch radii $r=20^\circ$ and $i=1$. The patch centers for which $\chi^\mathrm{2\,m}$, for the measured data, is higher than $\chi^\mathrm{2\,s}_1$ obtained from the randomly shuffled data and the significance $P^\mathrm{N}_1\leq 5$\,\% are shown in Fig.~\ref{fig:results_batse}.

The mean number of patches $\overline{N^\mathrm{s}_1}$ in the randomly shuffled data for which $\chi^\mathrm{2\,s}>\chi^\mathrm{2\,s}_{1}$ should be $\overline{N^\mathrm{s}_1}=10$ because we applied 1000 sky patches. The actual measured data gives $N^\mathrm{m}_1=28$. The corresponding chance probability of finding at least 28 patches on the sky with $\chi^\mathrm{2\,s}>\chi^\mathrm{2\,s}_{1}$ in the randomly shuffled data is $P^\mathrm{N}_1=4.1\,\%$.

There is no case of a patch, of a given radius and for a given tested GRB observable and given test statistic $\xi$ in the measured data which gives $\xi^\mathrm{m}$ higher than the highest $\xi^\mathrm{s}_{0}$ for patches of all random data shufflings.

In view of the fact that a large number of tests (eight different GRB observables, five different path radii and two different thresholds $i=1, 5$) were applied, one occurrence of $P^\mathrm{N}_\mathrm{i}<5$\,\% does not imply any significant anisotropy. Therefore, the results applied on the {\em CGRO}/BATSE sample are fully consistent with isotropy, as well.

\subsection{Results - {\em Swift}/BAT}
\label{sec:results_bat}

Tables~\ref{tab:results_bat_ks}--\ref{tab:results_bat_chi} summarize the results of all tested fluxes, fluences and duration of GRBs in the {\em CGRO}/BATSE sample using the Kolmogorov--Smirnov statistic $D$, Kuiper $V$, Anderson--Darling $AD$, and Chi-square $\chi^2$ statistics, respectively. Quantities in the columns have the same meaning as in the previous Tables~\ref{tab:results_gbm_ks}--\ref{tab:results_batse_chi}.

There were 18 cases with the chance probability $P^\mathrm{N}_\mathrm{i}\leq 5$\,\%, they are emphasized in boldface in the tables, and they were obtained for duration $T_{90}$ (patch radii $r=20^\circ$--$60^\circ$), fluence $S_1$ ($r=50^\circ$ and $60^\circ$) and peak flux $F_2$ ($r=40^\circ$ and $50^\circ$). These are the cases for which the statistical properties of GRBs are mostly deviated from the randomness. Fig.~\ref{fig:results_bat} shows the patch centers on the sky for which a given test statistic $\xi^\mathrm{m}$, for the measured data, is higher than $\xi^\mathrm{s}_\mathrm{i}$ obtained from the randomly shuffled data and the significance $P^\mathrm{N}_\mathrm{i}\leq 5$\,\%.

The lowest chance probability obtained was for Chi-square statistic, duration $T_{90}$, patch radii $r=40^\circ$, and threshold $i=5$. The mean number of patches $\overline{N^\mathrm{s}_5}$ in the randomly shuffled data for which $\chi^\mathrm{2\,s}>\chi^\mathrm{2\,s}_{5}$ should be $\overline{N^\mathrm{s}_5}=50$ because we applied 1000 sky patches. The actual measured data gives $N^\mathrm{m}_5=136$. The corresponding chance probability of finding at least 136 patches on the sky with $\chi^\mathrm{2\,s}>\chi^\mathrm{2\,s}_{5}$ in the randomly shuffled data is $P^\mathrm{N}_5=1.7\,\%$.

There is no case of a patch, of a given radius and for a given tested GRB observable and given test statistic $\xi$ in the measured data which gives $\xi^\mathrm{m}$ higher than the highest $\xi^\mathrm{s}_{0}$ for patches of all random data shufflings.

\section{Discussion}
\label{sec:discussion}

Table~\ref{tab:gbm_discussion} demonstrates how the signal of a feature found in work \citet{rip17} washed away with larger {\em Fermi}/GBM data sample used in this work and shows that the result do not point to any significant deviation from isotropy.

In case of the results obtained for the {\em Swift}/BAT a deeper discussion is needed. Eighteen tests gave the chance probability $P^\mathrm{N}_\mathrm{i}\leq 5$\,\%. Since we performed 600 statistical tests (15 tested GRB observables, 5 patch radii, 4 test statistics, and 2 limiting thresholds) it must be considered that the number of trials is high. However, not all of the tests are independent.

If we assume that there are $n=600$ independent tests (trials), with the probability of success in a single test $p$, one can calculate the probability ($p-value$) of finding at least $m$ successes in all trials from the binomial test. For details see for example Eq. (1) and (2) of \citet{vav08}. From Tables~\ref{tab:results_bat_ks} -- \ref{tab:results_bat_chi} one can see that we obtained $m=18$ for $p \leq 5$\,\%, $m=9$ for $p \leq 3$\,\%, and $m=4$ for $p \leq 2$\,\%. If we assume that $p=5$\,\% then we obtain for $m=18$, $m=9$, and $m=4$ the following probabilities: $p-value=99.4$\,\%, $p-value=99.4$\,\%, and $p-value=99.8$\,\%, respectively.

This is a coarse estimation because the tests are not independent. The number of independent tests (trials) is likely tens.

We obtained that in one test the lowest chance probability obtained was $P^\mathrm{N}_{5}=1.7$\,\% (Chi-square statistic, duration $T_{90}$, and patch radii $r=40^\circ$). However, for example, if the number of independent tests is 10 then for $p=1.7$\,\% the $p-value$ of the binomial test is 16\,\%. If the number of independent tests is 20 then the $p-value=29$\,\%. This suggests that a chance probability in a single test of 1.7\,\% is actually much more likely to occur by chance in one of the many tests used.

In order to estimate more precisely the chance probability of obtaining certain number of cases with $P^\mathrm{N}_\mathrm{i}$ less or equal a given limit, we performed 20 Monte Carlo (MC) simulations. In each simulation we randomly shuffled the values of each observable in the measured database and then run the whole analysis with this shuffled database instead of the actual measured database. We obtained that in 17 out of 20 MC simulations there were $m \ge 18$ number of tests giving $P^\mathrm{N}_\mathrm{i}\leq 5$\,\%. Therefore one can say that there is approximately 85\,\% of chance probability of obtaining at least 18 times the probability $P^\mathrm{N}_\mathrm{i}\leq 5$\,\% among the performed 600 tests in the isotropic randomized sample. We obtained the same percentage when we looked at the $m \ge 9$ number of tests giving $P^\mathrm{N}_\mathrm{i}\leq 3$\,\% and when we looked at the $m \ge 4$ number of tests giving $P^\mathrm{N}_\mathrm{i}\leq 2$\,\%. This is a key argument suggesting that also the results obtained for the {\em Swift}/BAT data sample are consistent with isotropy.

\begin{figure*}
\begin{tabular}{cc}
\includegraphics[width=0.45\textwidth]{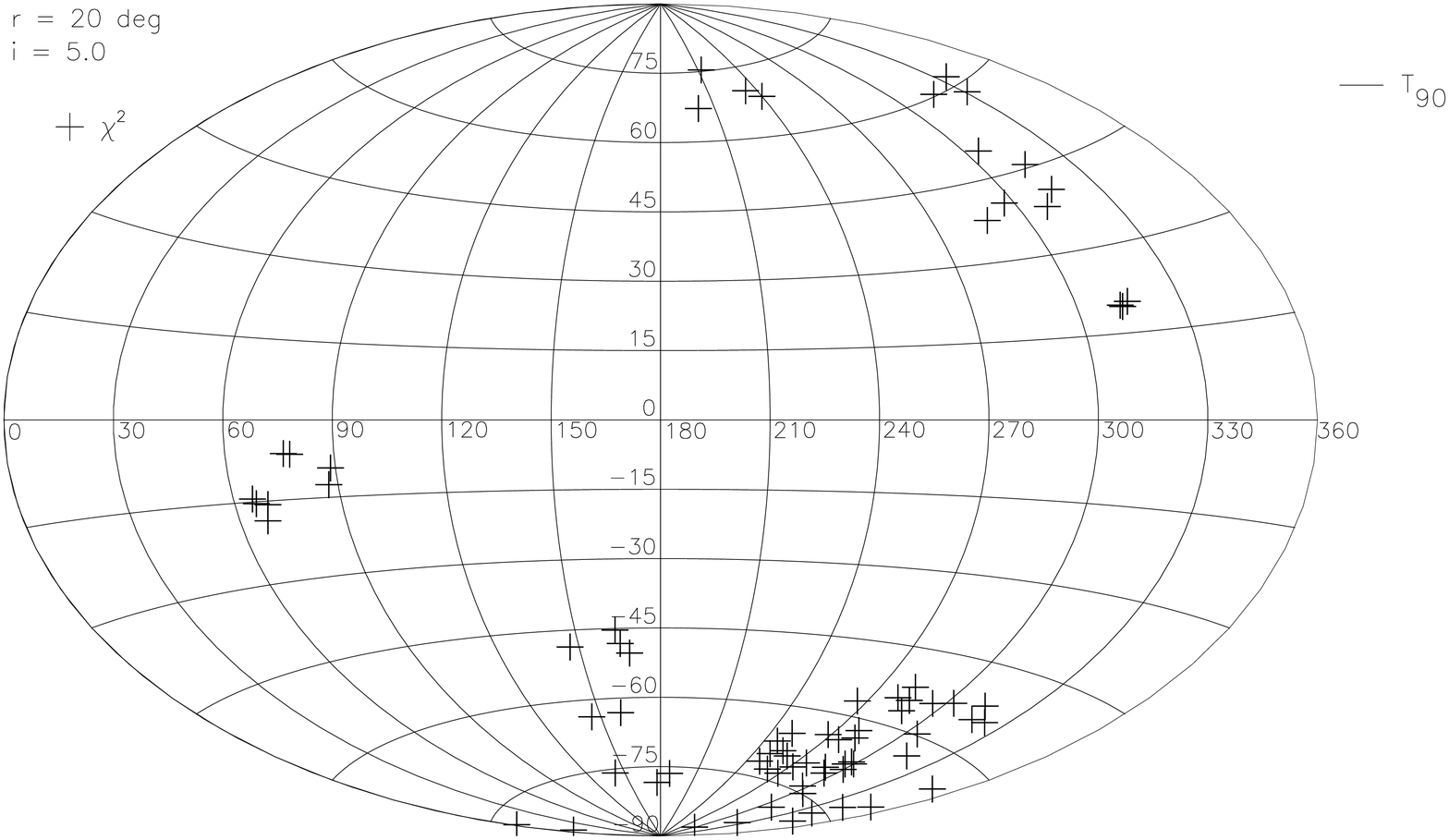} & \includegraphics[width=0.45\textwidth]{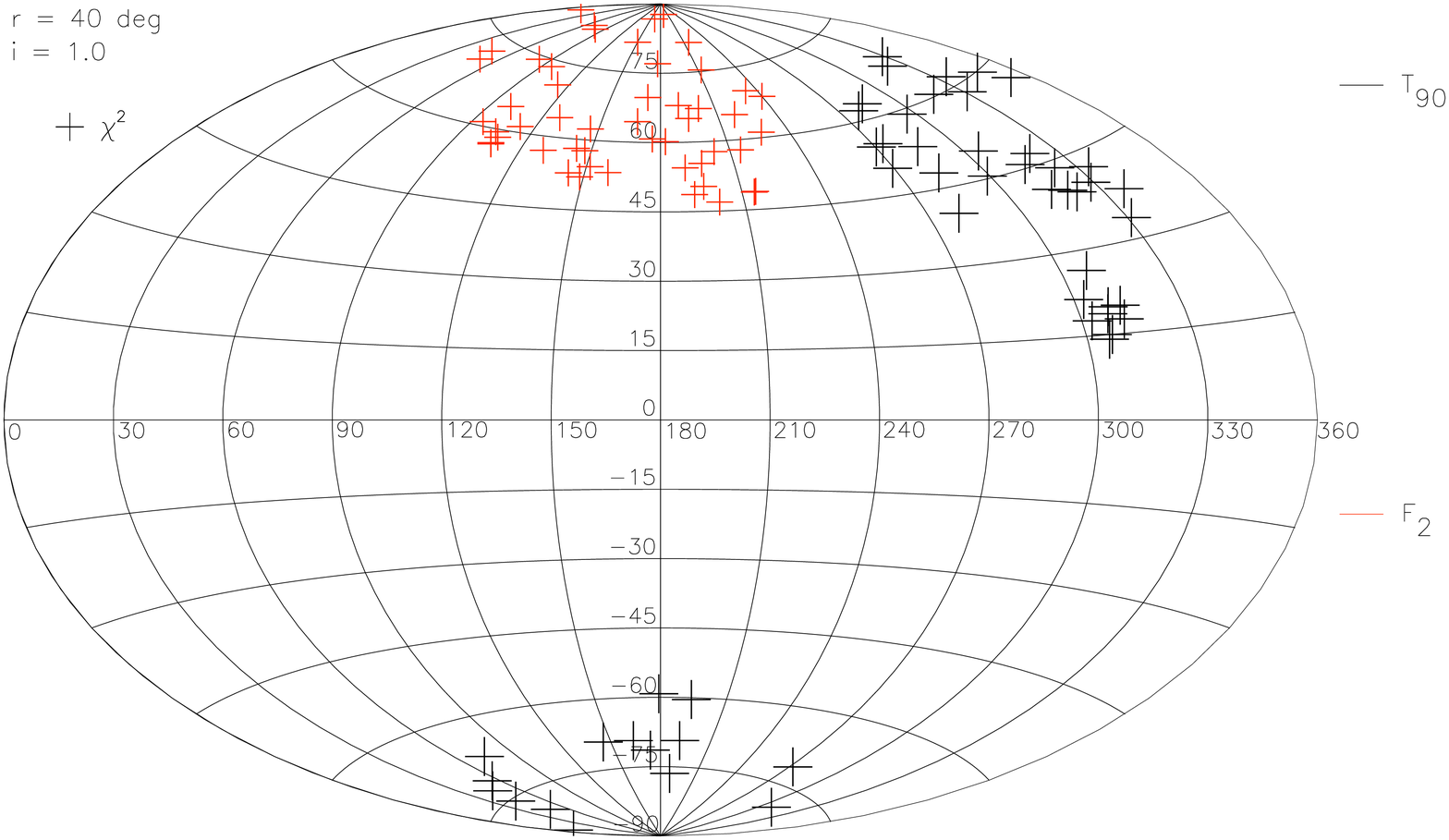} \\ [1.0ex]
\includegraphics[width=0.45\textwidth]{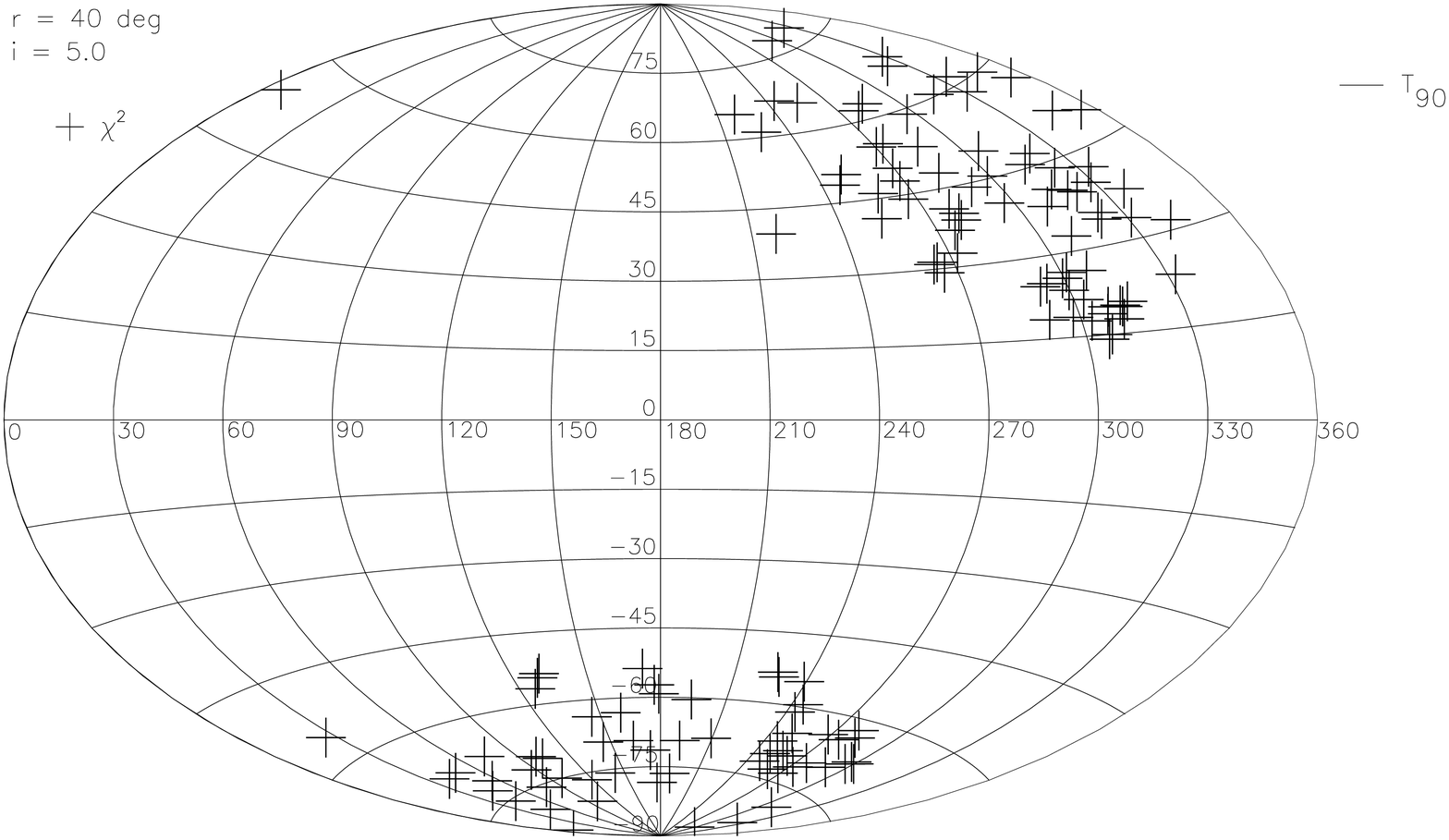} & \includegraphics[width=0.45\textwidth]{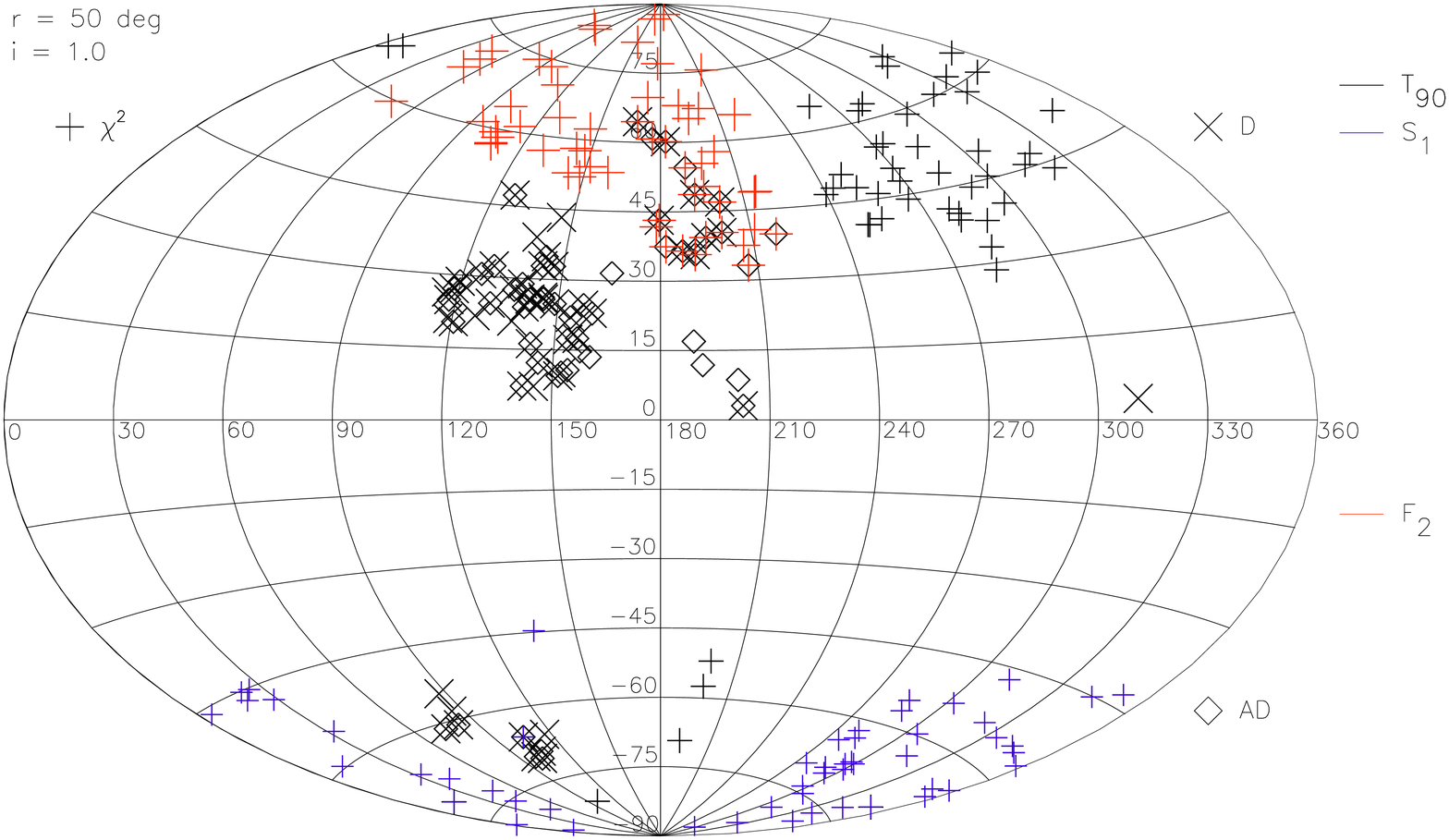}\\ [1.0ex]
\includegraphics[width=0.45\textwidth]{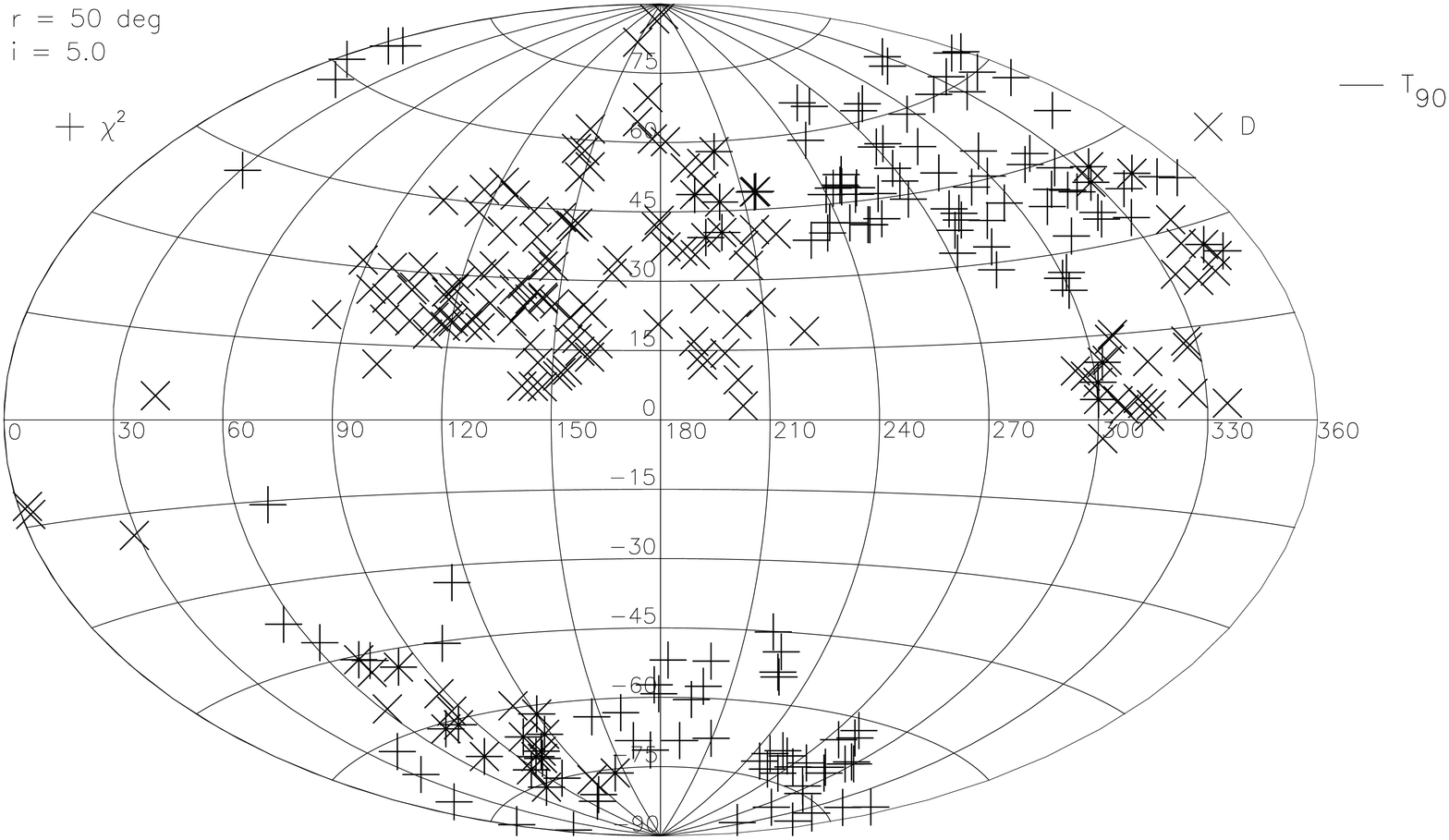} & \includegraphics[width=0.45\textwidth]{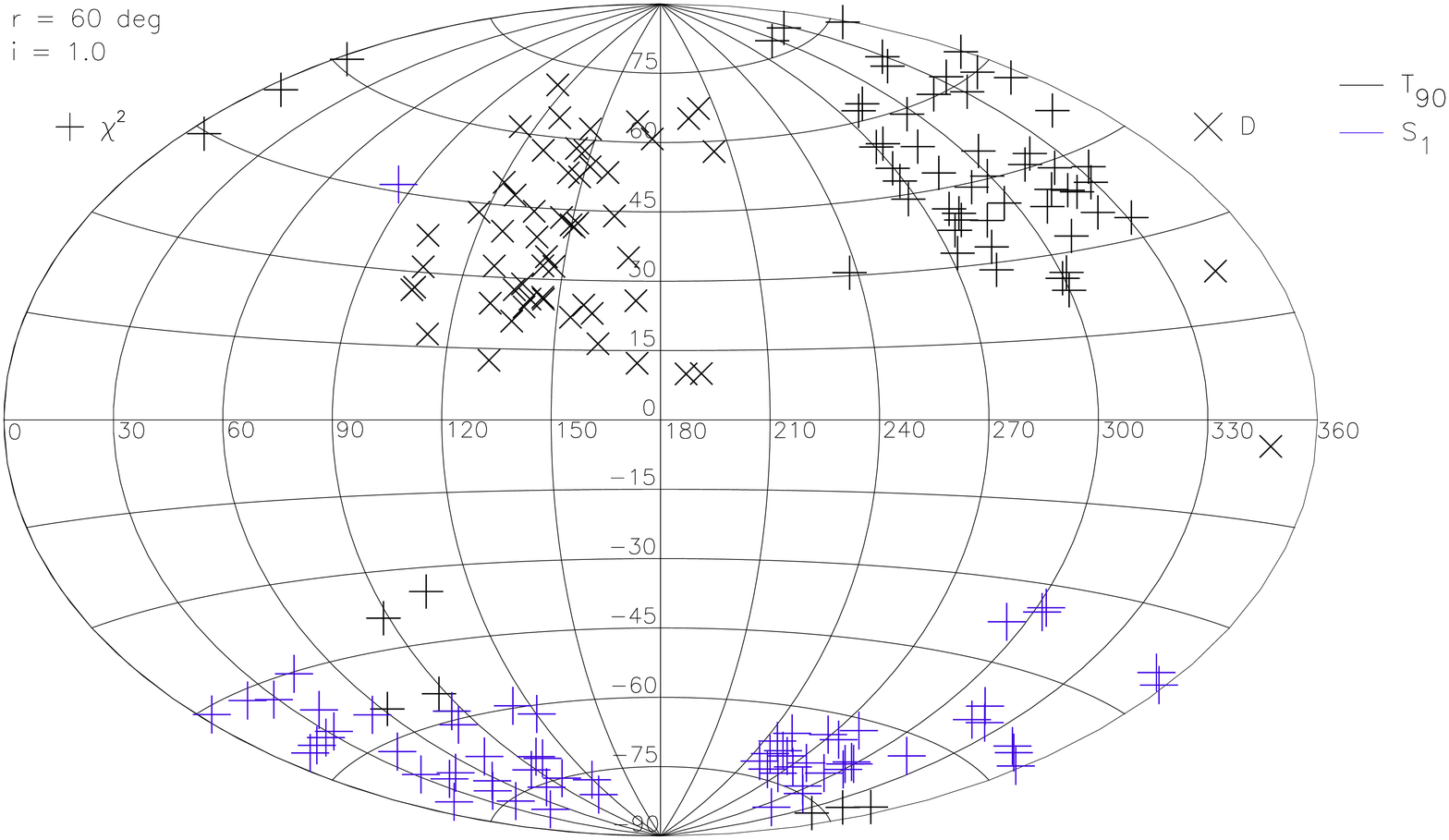}\\ [1.0ex]
\includegraphics[width=0.45\textwidth]{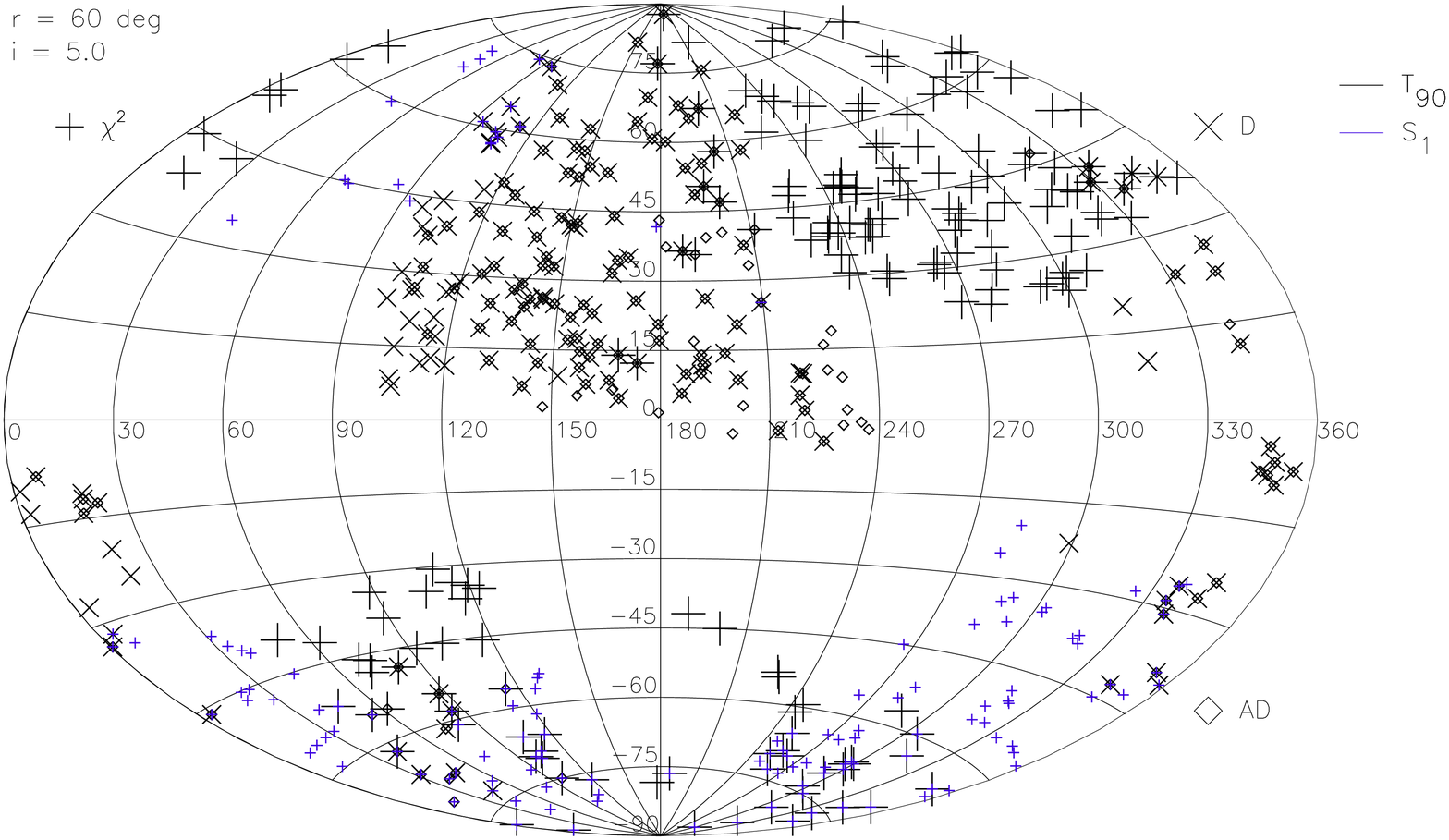} \\
\end{tabular}
\caption{Plotted are the patch centers on the sky in Galactic Coordinates (Aitoff projection), for which the statistical properties of GRBs are mostly deviated from the randomness in the {\em Swift}/BAT data sample. The combined results for different statistics $\xi=D, AD$, or $\chi^2$: Kolmogorov--Smirnov's $D$ (crosses), Anderson--Darling's $AD$ (diamonds), or Chi-square $\chi^2$ (pluses) are marked. The markers denote the centers of the patches for which a given statistic $\xi^\mathrm{m}$, for the measured data, is higher than $\xi^\mathrm{s}_\mathrm{i}$ obtained from the randomly shuffled data and the significance $P^\mathrm{N}_\mathrm{i}\leq 5$\,\%, where i=5 or 1. The size of the markers is inverse proportional to the probability $P^\mathrm{N}_\mathrm{i}$. Different colors mean different properties of GRBs being tested. In different panels the results are plotted separately for different patch radii $r$ and for different $i$.
}
\label{fig:results_bat}
\end{figure*}

Fig.~\ref{fig:results_bat} shows the patch centres for which a given statistic $\xi^\mathrm{m}$, for the measured {\em Swift}/BAT data, is higher than $\xi^\mathrm{s}_\mathrm{i}$ obtained from the randomly shuffled data and the significance $P^\mathrm{N}_\mathrm{i}\leq 5$\,\%, where i=5 or 1. From that figure it can appear that because of the clustering of these patch centers these deviations might be real or due to systematic differences in how the distributions have been sampled. However, nearby patches do not contain independent samples because they have rather large radii and they overlap. If there are randomly few GRBs with, for example, very high or low flux close to each other then a bundle of nearby patches will contain these GRBs and the distributions of fluxes in all these patches will be affected. It is important to mention that similar ``clustering" is seen also for the results of the MC simulations mentioned in the previous paragraph, where the whole analysis was run with randomly shuffled GRB properties in the catalog.

While discussing the results it should be mentioned that some selection effects could effect the results, for example the background variation on the sky or adjustment of the on-board trigger setting throughout the mission. On the other hand, since we found results consistent with isotropy possible selection effects would need to cancel out potential anisotropic features in the GRB properties.

\begin{table}
\centering
\caption{Comparison of some most significant results of several tests performed on the old (1591 GRBs in \citet{rip17}) and the new (2266 GRBs in this work) data samples of {\em Fermi}/GBM for Kolmogorov--Smirnov statistic $D$ and patch radii $r=20^\circ$ demonstrating how the signal of a feature found in work \citet{rip17} washed away with larger data sample.}
\label{tab:gbm_discussion}
\begin{tabular}{lllll|llll}
\hline
               &  & \multicolumn{2}{c}{old sample} &  & \multicolumn{4}{c}{new sample}   \\[0.5ex]
               & $N^\mathrm{m}_{5}$ & $P^\mathrm{N}_{5}$ & $N^\mathrm{m}_{1}$ & $P^\mathrm{N}_{1}$  & $N^\mathrm{m}_{5}$ & $P^\mathrm{N}_{5}$& $N^\mathrm{m}_{1}$ & $P^\mathrm{N}_{1}$ \\[0.5ex]
               &    & (\%) &    & (\%) &    & (\%) &    & (\%) \\
\hline
$F_{64}$       & 87 & 3.4  & 30 & 1.5  & 55 & 36.8 & 10 & 44.6 \\
$S$            & 72 & 13.1 & 30 & 2.2  & 57 & 32.2 & 17 & 17.1 \\
$S_\mathrm{B}$ & 74 & 11.5 & 31 & 1.4  & 52 & 43.7 & 17 & 17.9 \\
\hline
\end{tabular}
\end{table}

\section{Conclusions}
\label{sec:conclusion}

We inspected the isotropy of the observed properties of GRBs (not the distribution of their number density on the sky) such as their duration, fluences and peak fluxes at various energy bands and different time scales by a novel method - in its original form proposed in our previous work - applied on three major GRB catalogs. The whole GRB samples containing bursts regardless of their duration, spectral properties, luminosity or measured redshift, were used. The conclusions are following:

\begin{enumerate}
\item \noindent\parbox[t]{0.92\columnwidth}{We slightly improved the sensitivity of the original method proposed in work \citet{rip17}. Contrary to the original method, where the distributions of a given measured GRB property from large number of randomly spread patches on the sky were compared with a distribution of the same GRB property for the whole sky, herein we compare the distributions from the patches with the distributions obtained from the complement area of that patches.}
\\
\item \noindent\parbox[t]{0.92\columnwidth}{The method was applied to a considerably updated whole {\em Fermi}/GBM data sample containing 2266 GRBs, thus $\sim 40$\,\% larger than the sample investigated in work \citet{rip17}. The signal of a feature found in work \citet{rip17} washed away with larger data sample and the results are consistent with isotropy confirming our previous conclusions.}
\\
\item \noindent\parbox[t]{0.92\columnwidth}{The method was also applied to another large GRB dataset, the BATSE Current GRB Catalog of the {\em CGRO} satellite, which contains $\sim 2000$ bursts. The results based on the whole samples are consistent with isotropy as well.}
\\
\item \noindent\parbox[t]{0.92\columnwidth}{The last investigated whole GRB data sample is the {\em Swift}/BAT Gamma-Ray Burst Catalog containing $\sim 1200$ bursts. The localization accuracy of the {\em Swift}/BAT instrument is significantly better, $\sim 2$ arcmin compared to few degrees of {\em Fermi}/GBM or {\em CGRO}/BATSE. Therefore investigation of this sample is beneficial, too. The results are also consistent with isotropy.}
\end{enumerate}

\section*{Acknowledgements}

We acknowledge use of the {\em Fermi}/GBM, {\em CGRO}/BATSE, and {\em Swift}/BAT data. This research has made use of data, software and/or web tools obtained from the High Energy Astrophysics Science Archive Research Center (HEASARC), a service of the Astrophysics Science Division at NASA/GSFC and of the Smithsonian Astrophysical Observatory's High Energy Astrophysics Division. This work made use of procedures {\em KSTWO} and {\em KUIPERTWO} of the IDL software and its Astronomy Users library, and the package {\em adk\_1.0-2} of the R software. We kindly thank to A. M{\'e}sz{\'a}ros and M. Tarnopolski for useful comments and suggestions. J.R. is supported by the Lend\"ulet LP2016-11 grant awarded by the Hungarian Academy of Sciences. The research has been supported by the European Union, co-financed by the European Social Fund (Research and development activities at the E\"{o}tv\"{o}s Lor\'{a}nd University's Campus in Szombathely, EFOP-3.6.1-16-2016-00023). A.S. would like to acknowledge the support of the National Research Foundation of Korea (NRF-2016R1C1B2016478). A.S. would like to acknowledge the support of the Korea Institute for Advanced Study (KIAS) grant funded by the Korea government.

\clearpage

\begin{table}
%\scriptsize
\footnotesize
\caption{Results using the Kolmogorov--Smirnov statistic $D$ for the {\em Fermi}/GBM sample.}
\label{tab:results_gbm_ks}
\setlength\tabcolsep{4.0pt} % default value: 6pt
\begin{tabular}{llllllllll}
\hline	\\[-2.5ex]
Obs.	&	$r^\ast$	&	$D^\mathrm{s}_{5}$$^\dagger$	&	$N^\mathrm{m}_{5}$$^\ddagger$	&	$P^\mathrm{N}_{5}$$^\#$	&	$D^\mathrm{s}_{1}$$^\dagger$	&	$N^\mathrm{m}_{1}$$^\ddagger$	&	$P^\mathrm{N}_{1}$$^\#$	&	$D^\mathrm{s}_0$$^\dagger$	&	$D^\mathrm{m}_0$$^\S$	\\[0.5ex]
	&	($^\circ$)	&		&		&	(\%)	&		&		&	(\%)	&		&		\\[0.5ex]
\hline
$T_{90}$	&	20	&	0.17	&	70	&	15.8	&	0.20	&	13	&	29.4	&	0.36	&	0.27	\\[-0.3ex]
$T_{90}$	&	30	&	0.11	&	66	&	25.9	&	0.14	&	10	&	37.7	&	0.22	&	0.18	\\[-0.3ex]
$T_{90}$	&	40	&	0.09	&	78	&	20.5	&	0.11	&	22	&	15.0	&	0.18	&	0.12	\\[-0.3ex]
$T_{90}$	&	50	&	0.07	&	95	&	14.4	&	0.09	&	11	&	26.9	&	0.14	&	0.10	\\[-0.3ex]
$T_{90}$	&	60	&	0.07	&	93	&	16.8	&	0.08	&	4	&	42.4	&	0.12	&	0.08	\\
\end{tabular}
\end{table}

%\clearpage

\begin{table}
%\scriptsize
\footnotesize
\caption{Results using the Kuiper statistic $V$ for the {\em Fermi}/GBM sample.}
\label{tab:results_gbm_kuiper}
\setlength\tabcolsep{4.0pt} % default value: 6pt
\begin{tabular}{llllllllll}
\hline	\\[-2.5ex]
Obs.	&	$r^\ast$	&	$V^\mathrm{s}_{5}$$^\dagger$	&	$N^\mathrm{m}_{5}$$^\ddagger$	&	$P^\mathrm{N}_{5}$$^\#$	&	$V^\mathrm{s}_{1}$$^\dagger$	&	$N^\mathrm{m}_{1}$$^\ddagger$	&	$P^\mathrm{N}_{1}$$^\#$	&	$V^\mathrm{s}_0$$^\dagger$	&	$V^\mathrm{m}_0$$^\S$	\\[0.5ex]
	&	($^\circ$)	&		&		&	(\%)	&		&		&	(\%)	&		&		\\[0.5ex]
\hline
$T_{90}$	&	20	&	0.21	&	70	&	14.4	&	0.25	&	19	&	11.8	&	0.38	&	0.31	\\[-0.3ex]
$T_{90}$	&	30	&	0.15	&	88	&	9.2	&	0.17	&	21	&	13.1	&	0.26	&	0.20	\\[-0.3ex]
$T_{90}$	&	40	&	0.11	&	98	&	11.2	&	0.13	&	15	&	24.7	&	0.20	&	0.14	\\[-0.3ex]
$T_{90}$	&	50	&	0.09	&	114	&	8.8	&	0.11	&	16	&	20.5	&	0.16	&	0.12	\\[-0.3ex]
$T_{90}$	&	60	&	0.08	&	86	&	18.3	&	0.10	&	6	&	38.0	&	0.14	&	0.10	\\
\end{tabular}
\end{table}

\clearpage

\begin{table}
%\scriptsize
\footnotesize
\caption{Results using the Anderson--Darling statistic $AD$ for the {\em Fermi}/GBM sample.}
\label{tab:results_gbm_ad}
\setlength\tabcolsep{3.0pt} % default value: 6pt
\begin{tabular}{llllllllll}
\hline	\\[-2.5ex]
Obs.	&	$r^\ast$	&	$AD^\mathrm{s}_{5}$$^\dagger$	&	$N^\mathrm{m}_{5}$$^\ddagger$	&	$P^\mathrm{N}_{5}$$^\#$	&	$AD^\mathrm{s}_{1}$$^\dagger$	&	$N^\mathrm{m}_{1}$$^\ddagger$	&	$P^\mathrm{N}_{1}$$^\#$	&	$AD^\mathrm{s}_0$$^\dagger$	&	$AD^\mathrm{m}_0$$^\S$	\\[0.5ex]
	&	($^\circ$)	&		&		&	(\%)	&		&		&	(\%)	&		&		\\[0.5ex]
\hline
$T_{90}$	&	20	&	2.51	&	59	&	32.0	&	3.91	&	9	&	48.7	&	11.52	&	5.49	\\[-0.3ex]
$T_{90}$	&	30	&	2.50	&	70	&	24.3	&	3.91	&	17	&	21.4	&	12.35	&	5.97	\\[-0.3ex]
$T_{90}$	&	40	&	2.49	&	101	&	10.9	&	3.86	&	18	&	19.7	&	10.91	&	5.15	\\[-0.3ex]
$T_{90}$	&	50	&	2.50	&	110	&	10.8	&	3.90	&	28	&	11.5	&	11.87	&	6.39	\\[-0.3ex]
$T_{90}$	&	60	&	2.46	&	116	&	11.6	&	3.84	&	11	&	25.0	&	10.83	&	5.65	\\
\end{tabular}
\end{table}

%\clearpage

\begin{table}
%\scriptsize
\footnotesize
\caption{Results using the $\chi^2$ statistic for the {\em Fermi}/GBM sample.}
\label{tab:results_gbm_chi}
\setlength\tabcolsep{3.0pt} % default value: 6pt
\begin{tabular}{llllllllll}																	
\hline	\\[-2.5ex]
Obs.	&	$r^\ast$	&	$\chi^\mathrm{2\,s}_{5}$$^\dagger$	&	$N^\mathrm{m}_{5}$$^\ddagger$	&	$P^\mathrm{N}_{5}$$^\#$	&	$\chi^\mathrm{2\,s}_{1}$$^\dagger$	&	$N^\mathrm{m}_{1}$$^\ddagger$	&	$P^\mathrm{N}_{1}$$^\#$	&	$\chi^\mathrm{2\,s}_0$$^\dagger$	&	$\chi^\mathrm{2\,m}_{0}$$^\S$	\\[0.5ex]
	&	($^\circ$)	&		&		&	(\%)	&		&		&	(\%)	&		&		\\[0.5ex]
\hline
$T_{90}$	&	20	&	20.49	&	32	&	95.7	&	42.28	&	13	&	32.1	&	105.65	&	50.59	\\[-0.3ex]
$T_{90}$	&	30	&	20.70	&	69	&	16.9	&	27.35	&	17	&	21.9	&	72.21	&	32.31	\\[-0.3ex]
$T_{90}$	&	40	&	17.99	&	77	&	18.5	&	23.39	&	23	&	14.0	&	64.98	&	28.85	\\[-0.3ex]
$T_{90}$	&	50	&	17.04	&	85	&	17.0	&	21.97	&	10	&	31.6	&	47.40	&	26.14	\\[-0.3ex]
$T_{90}$	&	60	&	16.66	&	56	&	34.6	&	21.22	&	4	&	46.6	&	40.71	&	23.59	\\
\end{tabular}
\end{table}

\clearpage

\begin{table}
%\scriptsize
\footnotesize
\caption{Results using the Kolmogorov--Smirnov statistic $D$ for the {\em CGRO}/BATSE sample.}
\label{tab:results_batse_ks}
\setlength\tabcolsep{3.5pt} % default value: 6pt
\begin{tabular}{llllllllll}
\hline	\\[-2.5ex]
Obs.	&	$r^\ast$	&	$D^\mathrm{s}_{5}$$^\dagger$	&	$N^\mathrm{m}_{5}$$^\ddagger$	&	$P^\mathrm{N}_{5}$$^\#$	&	$D^\mathrm{s}_{1}$$^\dagger$	&	$N^\mathrm{m}_{1}$$^\ddagger$	&	$P^\mathrm{N}_{1}$$^\#$	&	$D^\mathrm{s}_0$$^\dagger$	&	$D^\mathrm{m}_0$$^\S$	\\[0.5ex]
	&	($^\circ$)	&		&		&	(\%)	&		&		&	(\%)	&		&		\\[0.5ex]
\hline
$T_{90}$	&	20	&	0.18	&	48	&	49.4	&	0.21	&	3	&	85.1	&	0.44	&	0.25	\\[-0.3ex]
$T_{90}$	&	30	&	0.12	&	29	&	77.1	&	0.14	&	8	&	45.2	&	0.23	&	0.17	\\[-0.3ex]
$T_{90}$	&	40	&	0.09	&	23	&	74.4	&	0.11	&	2	&	71.6	&	0.17	&	0.12	\\[-0.3ex]
$T_{90}$	&	50	&	0.08	&	28	&	60.6	&	0.09	&	5	&	44.0	&	0.17	&	0.10	\\[-0.3ex]
$T_{90}$	&	60	&	0.07	&	11	&	74.1	&	0.08	&	0	&	100.0	&	0.13	&	0.07	\\
\end{tabular}
\end{table}

%\clearpage

\begin{table}
%\scriptsize
\footnotesize
\caption{Results using the Kuiper statistic $V$ for the {\em CGRO}/BATSE sample.}
\label{tab:results_batse_kuiper}
\setlength\tabcolsep{4.0pt} % default value: 6pt
\begin{tabular}{llllllllll}
\hline	\\[-2.5ex]
Obs.	&	$r^\ast$	&	$V^\mathrm{s}_{5}$$^\dagger$	&	$N^\mathrm{m}_{5}$$^\ddagger$	&	$P^\mathrm{N}_{5}$$^\#$	&	$V^\mathrm{s}_{1}$$^\dagger$	&	$N^\mathrm{m}_{1}$$^\ddagger$	&	$P^\mathrm{N}_{1}$$^\#$	&	$V^\mathrm{s}_0$$^\dagger$	&	$V^\mathrm{m}_0$$^\S$	\\[0.5ex]
	&	($^\circ$)	&		&		&	(\%)	&		&		&	(\%)	&		&		\\[0.5ex]
\hline
$T_{90}$	&	20	&	0.23	&	26	&	92.9	&	0.26	&	3	&	89.2	&	0.49	&	0.28	\\[-0.3ex]
$T_{90}$	&	30	&	0.15	&	13	&	96.8	&	0.18	&	1	&	92.7	&	0.27	&	0.18	\\[-0.3ex]
$T_{90}$	&	40	&	0.12	&	1	&	99.6	&	0.14	&	0	&	100.0	&	0.21	&	0.12	\\[-0.3ex]
$T_{90}$	&	50	&	0.10	&	1	&	98.5	&	0.12	&	0	&	100.0	&	0.17	&	0.10	\\[-0.3ex]
$T_{90}$	&	60	&	0.09	&	3	&	93.0	&	0.10	&	0	&	100.0	&	0.15	&	0.10	\\
\end{tabular}
\end{table}

\clearpage

\begin{table}
%\scriptsize
\footnotesize
\caption{Results using the Anderson--Darling statistic $AD$ for the {\em CGRO}/BATSE sample.}
\label{tab:results_batse_ad}
\setlength\tabcolsep{3.0pt} % default value: 6pt
\begin{tabular}{llllllllll}
\hline	\\[-2.5ex]
Obs.	&	$r^\ast$	&	$AD^\mathrm{s}_{5}$$^\dagger$	&	$N^\mathrm{m}_{5}$$^\ddagger$	&	$P^\mathrm{N}_{5}$$^\#$	&	$AD^\mathrm{s}_{1}$$^\dagger$	&	$N^\mathrm{m}_{1}$$^\ddagger$	&	$P^\mathrm{N}_{1}$$^\#$	&	$AD^\mathrm{s}_0$$^\dagger$	&	$AD^\mathrm{m}_0$$^\S$	\\[0.5ex]
	&	($^\circ$)	&		&		&	(\%)	&		&		&	(\%)	&		&		\\[0.5ex]
\hline
$T_{90}$	&	20	&	2.51	&	49	&	47.2	&	3.90	&	15	&	23.8	&	11.95	&	5.34	\\[-0.3ex]
$T_{90}$	&	30	&	2.48	&	53	&	40.0	&	3.85	&	10	&	37.8	&	13.49	&	6.08	\\[-0.3ex]
$T_{90}$	&	40	&	2.52	&	62	&	30.3	&	3.91	&	12	&	26.9	&	12.31	&	6.71	\\[-0.3ex]
$T_{90}$	&	50	&	2.49	&	52	&	36.1	&	3.84	&	19	&	18.1	&	11.69	&	5.66	\\[-0.3ex]
$T_{90}$	&	60	&	2.50	&	48	&	37.7	&	3.89	&	2	&	48.4	&	14.28	&	4.01	\\
\end{tabular}
\end{table}

%\clearpage

\begin{table}
%\scriptsize
\footnotesize
\caption{Results using the $\chi^2$ statistic for the {\em CGRO}/BATSE sample.}
\label{tab:results_batse_chi}
\setlength\tabcolsep{3.0pt} % default value: 6pt
\begin{tabular}{llllllllll}
\hline	\\[-2.5ex]
Obs.	&	$r^\ast$	&	$\chi^\mathrm{2\,s}_{5}$$^\dagger$	&	$N^\mathrm{m}_{5}$$^\ddagger$	&	$P^\mathrm{N}_{5}$$^\#$	&	$\chi^\mathrm{2\,s}_{1}$$^\dagger$	&	$N^\mathrm{m}_{1}$$^\ddagger$	&	$P^\mathrm{N}_{1}$$^\#$	&	$\chi^\mathrm{2\,s}_0$$^\dagger$	&	$\chi^\mathrm{2\,m}_{0}$$^\S$	\\[0.5ex]
	&	($^\circ$)	&		&		&	(\%)	&		&		&	(\%)	&		&		\\[0.5ex]
\hline
$T_{90}$	&	20	&	19.96	&	49	&	50.2	&	42.36	&	2	&	85.1	&	88.46	&	47.29	\\[-0.3ex]
$T_{90}$	&	30	&	20.51	&	39	&	70.1	&	26.81	&	0	&	100.0	&	61.83	&	26.57	\\[-0.3ex]
$T_{90}$	&	40	&	17.96	&	32	&	67.7	&	23.17	&	2	&	72.8	&	51.97	&	29.02	\\[-0.3ex]
$T_{90}$	&	50	&	16.91	&	55	&	36.6	&	21.72	&	9	&	35.8	&	51.18	&	26.94	\\[-0.3ex]
$T_{90}$	&	60	&	16.84	&	53	&	37.6	&	21.54	&	11	&	28.1	&	44.06	&	26.19	\\
\end{tabular}
\end{table}

\clearpage

\begin{table}
\scriptsize
%\footnotesize
\caption{Results using the Kolmogorov--Smirnov statistic $D$ for the {\em Swift}/BAT sample.}
\label{tab:results_bat_ks}
\setlength\tabcolsep{4.0pt} % default value: 6pt
\begin{tabular}{llllllllll}
\hline	\\[-2.5ex]

Obs.	&	$r$	&	$D^\mathrm{s}_{5}$	&	$N^\mathrm{m}_{5}$	&	$P^\mathrm{N}_{5}$	&	$D^\mathrm{s}_{1}$	&	$N^\mathrm{m}_{1}$	&	$P^\mathrm{N}_{1}$	&	$D^\mathrm{s}_0$	&	$D^\mathrm{m}_0$	\\[0.5ex]
	&	($^\circ$)	&		&		&	(\%)	&		&		&	(\%)	&		&		\\[0.5ex]
\hline
$T_{90}$         &      20  &        0.23   &        50   &             45.8     &        0.29   &        11   &             38.1     &        0.50  &      0.34   \\[-0.5ex]
$T_{90}$         &      30  &        0.16   &        68   &             22.5     &        0.19   &        12   &             32.3     &        0.32  &      0.22   \\[-0.5ex]
$T_{90}$         &      40  &        0.12   &       116   &              6.1     &        0.15   &        19   &             18.3     &        0.24  &      0.18   \\[-0.5ex]
$T_{90}$         &      50  &        0.10   &       163   &  {\bf        2.9  }  &        0.12   &        55   &  {\bf        2.9  }  &        0.19  &      0.16   \\[-0.5ex]
$T_{90}$         &      60  &        0.09   &       178   &  {\bf        4.0  }  &        0.11   &        56   &  {\bf        3.6  }  &        0.17  &      0.14   \\
\end{tabular}
\end{table}

%\clearpage

\begin{table}
\scriptsize
%\footnotesize
\caption{Results using the Kuiper statistic $V$ for the {\em Swift}/BAT sample.}
\label{tab:results_bat_kuiper}
\setlength\tabcolsep{4.0pt} % default value: 6pt
\begin{tabular}{llllllllll}
\hline	\\[-2.5ex]
Obs.	&	$r$	&	$V^\mathrm{s}_{5}$	&	$N^\mathrm{m}_{5}$	&	$P^\mathrm{N}_{5}$	&	$V^\mathrm{s}_{1}$	&	$N^\mathrm{m}_{1}$	&	$P^\mathrm{N}_{1}$	&	$V^\mathrm{s}_0$	&	$V^\mathrm{m}_0$	\\[0.5ex]
	&	($^\circ$)	&		&		&	(\%)	&		&		&	(\%)	&		&		\\[0.5ex]
\hline
$T_{90}$         &      20  &        0.30   &        45   &             59.3     &        0.35   &         9   &             50.9     &        0.54  &      0.37   \\[-0.5ex]
$T_{90}$         &      30  &        0.20   &        35   &             69.8     &        0.23   &         2   &             84.7     &        0.35  &      0.24   \\[-0.5ex]
$T_{90}$         &      40  &        0.16   &        42   &             51.8     &        0.18   &         2   &             72.5     &        0.27  &      0.20   \\[-0.5ex]
$T_{90}$         &      50  &        0.13   &        81   &             18.8     &        0.15   &        17   &             20.6     &        0.22  &      0.17   \\[-0.5ex]
$T_{90}$         &      60  &        0.12   &        68   &             25.4     &        0.13   &        16   &             19.6     &        0.19  &      0.14   \\
\end{tabular}
\end{table}

\clearpage

\begin{table}
\scriptsize
%\footnotesize
\caption{Results using the Anderson--Darling statistic $AD$ for the {\em Swift}/BAT sample.}
\label{tab:results_bat_ad}
\setlength\tabcolsep{3.0pt} % default value: 6pt
\begin{tabular}{llllllllll}
\hline	\\[-2.5ex]
Obs.	&	$r$	&	$AD^\mathrm{s}_{5}$	&	$N^\mathrm{m}_{5}$	&	$P^\mathrm{N}_{5}$	&	$AD^\mathrm{s}_{1}$	&	$N^\mathrm{m}_{1}$	&	$P^\mathrm{N}_{1}$	&	$AD^\mathrm{s}_0$	&	$AD^\mathrm{m}_0$	\\[0.5ex]
	&	($^\circ$)	&		&		&	(\%)	&		&		&	(\%)	&		&		\\[0.5ex]
\hline
$T_{90}$         &      20  &        2.50   &        61   &             26.6     &        3.90   &         8   &             52.0     &       12.96  &      4.80   \\[-0.5ex]
$T_{90}$         &      30  &        2.50   &        87   &             11.8     &        3.92   &        17   &             19.2     &       13.16  &      4.87   \\[-0.5ex]
$T_{90}$         &      40  &        2.49   &       115   &              6.5     &        3.86   &        25   &             13.3     &       12.67  &      5.55   \\[-0.5ex]
$T_{90}$         &      50  &        2.50   &       145   &              5.4     &        3.87   &        56   &  {\bf        3.5  }  &       10.71  &      7.04   \\[-0.5ex]
$T_{90}$         &      60  &        2.46   &       169   &  {\bf        4.9  }  &        3.84   &        44   &              6.3     &       10.53  &      6.97   \\
\end{tabular}
\end{table}

%\clearpage

\begin{table}
\scriptsize
%\footnotesize
\caption{Results using the $\chi^2$ statistic for the {\em Swift}/BAT sample.}
\label{tab:results_bat_chi}
\setlength\tabcolsep{3.0pt} % default value: 6pt
\begin{tabular}{llllllllll}																	
\hline	\\[-2.5ex]
Obs.	&	$r$	&	$\chi^\mathrm{2\,s}_{5}$	&	$N^\mathrm{m}_{5}$	&	$P^\mathrm{N}_{5}$	&	$\chi^\mathrm{2\,s}_{1}$	&	$N^\mathrm{m}_{1}$	&	$P^\mathrm{N}_{1}$	&	$\chi^\mathrm{2\,s}_0$	&	$\chi^\mathrm{2\,m}_{0}$	\\[0.5ex]
	&	($^\circ$)	&		&		&	(\%)	&		&		&	(\%)	&		&		\\[0.5ex]
\hline
$T_{90}$         &      20  &       21.09   &        77   &  {\bf        3.1  }  &       42.93   &         6   &             59.4     &      113.50  &     50.62   \\[-0.5ex]
$T_{90}$         &      30  &       20.97   &        62   &             30.0     &       28.02   &        25   &             10.7     &       67.82  &     41.37   \\[-0.5ex]
$T_{90}$         &      40  &       17.90   &       136   &  {\bf        1.7  }  &       23.25   &        52   &  {\bf        1.8  }  &       55.28  &     54.61   \\[-0.5ex]
$T_{90}$         &      50  &       17.07   &       155   &  {\bf        2.0  }  &       21.98   &        48   &  {\bf        3.3  }  &       47.86  &     40.90   \\[-0.5ex]
$T_{90}$         &      60  &       16.66   &       181   &  {\bf        2.3  }  &       21.27   &        61   &  {\bf        2.3  }  &       44.66  &     32.07   \\
\end{tabular}
\end{table}

\clearpage
%%%%%%%%%%%%%%%%%%%%%%%%%%%%%%%%%%%%%%%%%%%%%%%%%%


%%%%%%%%%%%%%%%%%%%%%%%%%%%%%%%%%%%%%%%%%%%%%%%%%%

%%%%%%%%%%%%%%%%% APPENDICES %%%%%%%%%%%%%%%%%%%%%

%\appendix

%%%%%%%%%%%%%%%%%%%%%%%%%%%%%%%%%%%%%%%%%%%%%%%%%%

% Don't change these lines
\bsp	% typesetting comment
\label{lastpage}
\end{document}